\documentclass[12pt]{article}

\usepackage{amssymb}
\usepackage{amsmath}

\usepackage{graphicx}

\usepackage{makecell}
\begin{document} %
%%%%%%%%%%%%%%%%%%

%%%%%%%%%%%%%%%%%%%%%%%%%%%%%%%%%%%%%%%%%%%%%%%%%%%%%%%%%%%%%%%%%%%%%

\title{Probing anomalous $Z\gamma\gamma\gamma$ couplings at a future muon collider}

\author{H. Amarkhail\thanks{Electronic address: hidayatamarkhail@gmail.com}
    \\
    {\small Department of Physics, Sivas Cumhuriyet University, 58140,
        Sivas, Turkey}
    \\
    {\small and Department of Physics, Kandahar University, Kandahar 3801, Afghanistan}
    \\
    S.C. \.{I}nan\thanks{Electronic address: sceminan@cumhuriyet.edu.tr}
    \\
    {\small Department of Physics, Sivas Cumhuriyet University, 58140,
        Sivas, Turkey}
    \\
    {\small and}
    \\
    A.V. Kisselev\thanks{Electronic address:
        alexandre.kisselev@ihep.ru} \\
    {\small A.A. Logunov Institute for High Energy Physics, NRC
        ``Kurchatov Institute'',}
    \\
    {\small 142281, Protvino, Russian Federation} }

\date{}

\maketitle

\begin{abstract}
The sensitivity to anomalous quartic gauge couplings  (AQGCs) of the
$\gamma\gamma\gamma Z$ interaction is studied in the $\mu^+\mu^-
\rightarrow \mu^+\gamma\gamma \mu^-$ scattering at a future muon
collider with unpolarized beams. The anomalous $\gamma\gamma\gamma
Z$ vertex is described by two couplings, $\zeta_1$ and $\zeta_2$.
The differential and total cross sections are calculated for the
center-of-mass energies of 3 TeV, 14 TeV, and 100 TeV. For these
values of the collision energy the $95\%$ C.L. exclusion regions for
AQGCs are obtained depending on the systematic error. In particular,
for the 14 TeV muon collider with the integrated luminosity $L = 20$
ab$^{-1}$ the best sensitivities are derived to be $\zeta_1 = 3.1
\times 10^{-5}$ TeV$^{-4}$ and $\zeta_2 = 6.5 \times 10^{-5}$
TeV$^{-4}$. These constraints are three orders of magnitude stronger
than the bounds obtained for the 27 TeV HE-LHC with $L = 15$
ab$^{-1}$. At the 100 TeV muon collider with $L = 1000$ ab$^{-1}$
AQGCs can be probed up to $(1.64 \div 3.4) \times 10^{-8}$
TeV$^{-4}$. The partial-wave unitarity constraints on couplings
$\zeta_1$, $\zeta_2$ are evaluated. It is shown that the unitarity
is not violated in the region of the AQGCs examined in the present
paper.
\end{abstract}

\maketitle

%%%%%%%%%%%%%%%%%%%%%%%%%%%%%%%%%%%%%%%%%%%%%%%%%%%%%%%%%%%%%%%%%%%%%

%%%%%%%%%%%%%%%%%%%%%%%%
\section{Introduction} %
%%%%%%%%%%%%%%%%%%%%%%%%

In the Standard Model (SM) there are only three quartic gauge
couplings which involve at least two charged $W$ bosons. In
particular, there is no contribution from the $Z\gamma\gamma\gamma$
coupling to the decay $Z \rightarrow \gamma\gamma\gamma$ which
proceeds exclusively via fermion and $W$-boson loops. First search
for anomalous couplings in $Z \rightarrow\gamma\gamma\gamma$ events
was made by the L3 collaboration at the LEP. The branching ratio
upper limit for the $Z$ boson decaying directly into three photons
was found to be $1.0 \times 10^{-5}$ \cite{L3:QGCs}. The LEP 2 data,
as was shown in \cite{Gutierrez:2014}, induce limits of order
$10^{-5}$ GeV$^{-4}$ for the anomalous $Z\gamma\gamma\gamma$
couplings and of order $10^{-3}$ GeV$^{-4}$ for the $ZZ\gamma\gamma$
couplings. The best upper bound for the $Z
\rightarrow\gamma\gamma\gamma$ decay has been obtained by the ATLAS
collaboration using $pp$ collisions at $\sqrt{s} = 8$ TeV and equals
to $\mathcal{B}(Z \rightarrow \gamma\gamma\gamma) < 2.2 \times
10^{-6}$ \cite{ATLAS:QGCs}. Note that the SM expectation is
$\mathcal{B}(Z \rightarrow \gamma\gamma\gamma) = 5.4 \times
10^{-10}$.

Analogously, there is no contribution from the $Z\gamma\gamma\gamma$
coupling to the $Z\gamma \rightarrow \gamma\gamma$ collision. Thus,
a diphoton production via VBF at $e^+e^-$ or muon colliders can be
regarded as a good process to probe effects induced by anomalous
quartic vertex $Z\gamma\gamma\gamma$.

The anomalous couplings for the $Z\gamma \rightarrow \gamma\gamma$
interaction is a particular case of the anomalous quartic gauge
couplings (AQGCs). They were searched for in a number of experiments
at the LHC, and upper bounds have been obtained
\cite{ATLAS:QGCs_1}-\cite{CMS_TOTEM:QGCs}. As for limits on AQGCs
for neutral bosons, they were derived both for the CLIC
\cite{{Koksal:2014}}-\cite{Gurkanli:2023_2} and LHC
\cite{{Eboli:2004}}-\cite{Yang:2021}. The upper bounds on AQGCs were
also obtained for future $\gamma e$ and $\gamma\gamma$ colliders
\cite{Eboli:1995}-\cite{Koksal:2016}, future hadron
\cite{Senol:2022_1}-\cite{Senol:2022}, and $eh$ colliders
\cite{Eboli:1994}-\cite{Gurkanli:2023_1}, see also
\cite{Stirling:2000}-\cite{Guo:2020_2}. Recently expected limits on
AQGCs at a muon collider have been examined
\cite{{I_K:2023}}-\cite{Zhang:2023}.

Previously, we studied the anomalous $\gamma\gamma\gamma Z$
couplings through $\gamma Z$ production in $\gamma\gamma$ collisions
at the CLIC~\cite{I_K:2021_2}. The goal of the present paper is to
probe these couplings in the diphoton production at a future muon
collider. The novelty of our work is that \emph{i}) we study the
muon collider instead of $e^+e^-$ collider; \emph{ii}) we consider a
lepton collision, while in \cite{I_K:2021_2} a $\gamma\gamma$ mode
of the $e^+e^-$ collider (namely, a collision of Compton
backscattered photons) was examined; \emph{iii}) we address
center-of-mass energies which are significantly higher than the CLIC
energies.

On the assumption that a new physics mass scale $\Lambda$ is larger
than the collision energy $\sqrt{s}$, all new physics manifestations
can be described using an effective Lagrangian
$\mathcal{L}_{\mathrm{EFT}}$ valid for $\sqrt{s} \gg \Lambda$ (see,
for instance, \cite{Gupta:2012}). For our goals, it is enough to use
dimension-8 operators
$B_{\mu\nu}B^{\nu\nu}B_{\rho\sigma}B^{\rho\varrho}$,
$W_{\mu\nu}W^{\nu\nu}W_{\rho\sigma}W^{\rho\varrho}$, and
$B_{\mu\nu}B^{\nu\nu}W_{\rho\sigma}W^{\rho\varrho}$, where $B_\mu$
and $W_\mu$ are the $U(1)_Y$ and $SU(2)$ gauge fields, respectively.
Note that the dual strength tensors are not considered. In a broken
phase (in which the Lagrangian is expressed in terms of physical
fields) a part of the effective Lagrangian describing anomalous
$Z\gamma\gamma\gamma$ interaction looks like \cite{Baldenegro:2017}
\begin{equation}\label{Lagrangian}
\mathcal{L}_{Z\gamma\gamma\gamma} = \zeta_1 O_1 + \zeta_2 O_2 \;,
\end{equation}
with the operators
\begin{equation}\label{operators_B}
O_1 =  F^{\mu\nu}F_{\mu\nu} F^{\rho\sigma}Z_{\rho\sigma} \;, \quad
O_2 = F^{\mu\nu} F_{\nu\rho}F^{\rho\sigma}Z_{\sigma\mu} \;,
\end{equation}
where $F_{\mu\nu} = \partial_\mu A_\nu - \partial_\nu A_\mu$, and
$Z_{\mu\nu} = \partial_\mu Z_\nu - \partial_\nu Z_\mu$. The
anomalous couplings $\zeta_1$ and $\zeta_2$ have dimension
TeV$^{-4}$. Instead of the  operator $O_2$ the
$F_{\mu\nu}\tilde{F}^{\mu\nu}F_{\rho\sigma} \tilde{Z}^{\rho\sigma}$
operator can be encountered, where $\tilde{F}^{\mu\nu} =
(1/2)\epsilon^{\mu\nu\alpha\beta}F_{\alpha\beta}$ and
$\tilde{Z}^{\mu\nu} =
(1/2)\epsilon^{\mu\nu\alpha\beta}Z_{\alpha\beta}$. The corresponding
Lagrangians can be transformed into each other (up to a full
derivative) using equations of motion.

As mentioned above, two operators in \eqref{Lagrangian} can
naturally arise from the effective Lagrangian
$\mathcal{L}_{\mathrm{EFT}}$. However, we may not appeal to
$\mathcal{L}_{\mathrm{EFT}}$, but just to demand that our operators
must be dimension-8 Lorentz invariant CP-even operators constructed
from one $Z$ boson and three photon fields. Then we come to the same
set of two operators \eqref{operators_B}. In such a case, anomalous
interactions \eqref{Lagrangian} can be considered independently of
$\mathcal{L}_{\mathrm{EFT}}$.

In a number of papers (see, for instance,
\cite{Koksal:2023}-\cite{Gurkanli:2023_2}) bounds on couplings
$f_i/\Lambda^4$ of an \emph{unbroken} Lagrangian (defined by the
fields $B_\mu$ and $W_\mu$) were searched for. In such an approach,
while obtaining bounds on any coupling, all other couplings are
assumed to be zero.

The anomalous couplings of the neutral gauge bosons in the
\emph{broken} phase (like our couplings $\zeta_1, \zeta_2$) are
known to be linear combinations of the couplings $f_i/\Lambda^4$. In
our approach, to examine the neutral AQGCs, we suggest that only one
of the quartic couplings, $\gamma\gamma\gamma\gamma$,
$Z\gamma\gamma\gamma$, or $ZZ\gamma\gamma\gamma$, is nonzero. It is
equivalent to an assumption that only one particular linear
combinations of $f_i/\Lambda^4$ is nonzero. We think that both
approaches have the right to exist.

In our previous work \cite{I_K:2023} we addressed anomalous
couplings of the $\gamma\gamma\gamma\gamma$ vertex. The goal of the
present paper is to study the couplings $\zeta_1, \zeta_2$ of the
anomalous $Z\gamma\gamma\gamma$ vertex \eqref{Lagrangian}.

%%%%%%%%%%%%%%%%%%%%%%%%%%%%%%%%%%%%%%%%%%%%%%%%%%%%%%%
\section{ Production of photon pair at muon collider} %
%%%%%%%%%%%%%%%%%%%%%%%%%%%%%%%%%%%%%%%%%%%%%%%%%%%%%%%

\begin{figure}[htb]
\begin{center}
\includegraphics[scale=0.5]{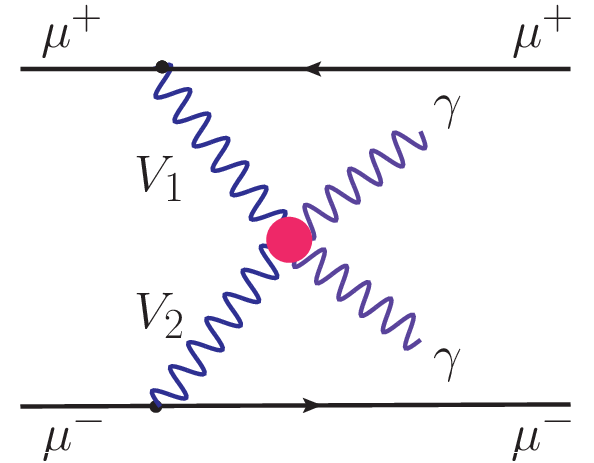}
\caption{The Feynman diagrams describing diphoton production in the
$\mu^+\mu^-$ collision via vector boson fusion.}
\label{fig:mu-VBF-mu_m}
\end{center}
\end{figure}

As is already mentioned in Introduction, our goal is to study AQGCs
in the diphoton production at the muon collider,
\begin{equation}\label{process}
\mu^+\mu^- \rightarrow \mu^+ V_1 V_2 \mu^- \rightarrow
\mu^+\gamma\gamma \mu^- ,
\end{equation}
which goes via VBF $V_1 V_2 \rightarrow \gamma\gamma$ ($V_{1,2} =
\gamma, Z$), see Fig.~\ref{fig:mu-VBF-mu_m}. The colliding muon
beams are assumed to be unpolarized, and we sum over the photon
polarization states. This process can be regarded as an
\emph{exclusive} process by requiring the outgoing muons to be
observable in the detector coverage
\begin{equation}\label{angle_cut}
10^\circ < \theta < 170^\circ \;,
\end{equation}
where $\theta$ is a scattering angle of the outgoing muons
\cite{Han:2023}.

%%%%%%%%%%%%%%%%%%%%%%%%%%%%%%%%%%%%%%%%
% Figure 1. Diphoton in muon collision %
%%%%%%%%%%%%%%%%%%%%%%%%%%%%%%%%%%%%%%%%

The cross section of the process \eqref{process} is
\begin{equation}\label{cs}
d\sigma = \int\limits_{\tau_{\min}}^{\tau_{\max}} \!\!d\tau
\!\!\!\int\limits_{x_{\min}}^{x_{\max}} \!\!\frac{dx}{x}
\,\sum_{V_1, V_2} \!\!f_{V_1/\mu^+}(x, Q^2) f_{V_2/\mu^-}(\tau/x,
Q^2) \,d\hat{\sigma} (V_1V_2\rightarrow \gamma\gamma) \;,
\end{equation}
where
\begin{equation}\label{y_z_limits}
x_{\max} = 1 - \frac{m_\mu}{E_\mu} \;, \ \tau_{\max} = \left( 1 -
\frac{m_\mu}{E_\mu} \right)^{\!2} , \ x_{\min} = \tau/x_{\max} \;, \
\tau_{\min} = \frac{p_\bot^2}{E_\mu^2} \;,
\end{equation}
$E_\mu$ is the energy of the muon beams ($4E_\mu^2 = s$), $p_\bot$
is the transverse momenta of the outgoing photons, and $m_\mu$ is
the muon mass. In \eqref{cs} $V_1$ and $V_2$ run over $\gamma, Z_+,
Z_-, Z_0$, where $Z_\pm$ ($Z_0$) denotes $Z$ boson with transverse
(longitudinal) helicity. The quantities $f_{V_{1,2}/\mu^\pm}(x,
Q^2)$ denote distributions of the gauge bosons in incoming muons.

For photon distributions we use the equivalent photon approximation
(EPA) \cite{Weizsacker:1934}-\cite{Carimalo:1979}. In the EPA the
polarized distributions of photon inside unpolarized fermion beam
are given by \cite{Budnev:1975}
\begin{align}\label{photon_spectrum}
f_{\gamma_\pm/f}(x, Q^2) = f_{\gamma/f}(x, Q^2) =
\frac{\alpha}{4\pi} \frac{1 + (1 - x)^2}{x}
\ln\frac{Q^2}{Q^2_{\min}} \;,
\end{align}
where $x = E_\gamma/E_\mu$ is fraction of the beam energy carried by
the photon, and $Q^2 = sx/2$. In our case $f = \mu^-$, $\bar{f} =
\mu^+$. Because of C and P invariance, $f_{\gamma/\bar{f}} =
f_{\gamma/f}$. For the $Z$ boson we use so-called effective $W$
approximation (EWA)~\cite{Dawson:1985}-\cite{Ruiz:2021}. It allows
to treat of massive vector bosons as partons inside the colliding
beams and reads \cite{Ruiz:2021}
\begin{align}
f_{Z_\pm/f}(x, Q^2) &= \frac{\alpha_Z}{4\pi} \frac{(g_V^f \mp
g_A^f)^2 + (g_V^f \pm g_A^f)^2(1 - x)^2}{x}
\ln\frac{Q^2}{Q^2_{\min}} \;,
\label{Z_trans} \\
f_{Z_0/f}(x, Q^2) &= \frac{\alpha_Z}{\pi} \frac{[(g_V^f)^2 +
(g_A^f)^2](1 - x)}{x} \;, \label{Z_long}
\end{align}
where
\begin{equation}\label{alpha_Z}
\alpha_Z = \frac{\alpha}{(\cos\theta_W \sin\theta_W)^2} \;, \ \
g_V^f = \frac{1}{2}(T_3^f)_L - Q^f \sin^2 \theta_W \;, \ \  g_A^f =
- \frac{1}{2}(T_3^f)_L  \;.
\end{equation}
There are CP relations $f_{Z_\pm/\bar{f}} = f_{Z_\mp/f}$,
$f_{Z_0/\bar{f}} = f_{Z_0/f}$.

Without constraints on the scattering angles $\theta$ of the
outgoing muons, $Q^2_{\min}$ has the following values
\begin{equation}\label{Q2_min_no_cuts}
Q^2_{\min} =
\left\{
       \begin{array}{ll}
       m_\mu^2 x^2/(1 - x), & \hbox{in eq.~\eqref{photon_spectrum};}
       \\
       m_Z^2, & \hbox{in eq.~\eqref{Z_trans}.}
       \end{array}
\right.
\end{equation}
For the exclusive diphoton production \eqref{process} with cuts
\eqref{angle_cut} one has to put in eqs.~\eqref{photon_spectrum},
\eqref{Z_trans} the quantity
\begin{equation}\label{Q2_min_with_cuts}
Q^2_{\min} = \frac{s(1-x)}{2} \!\left( \frac{1}{\cos\theta_{\min}} -
1 \right) ,
\end{equation}
where $\theta_{\min} = 10^\circ$.

In \eqref{cs} the VBF cross section is a sum of squared helicity
amplitudes,
\begin{equation}\label{subprocess_cs}
\frac{d\hat{\sigma}}{d\Omega}(V_1V_2\rightarrow \gamma\gamma) =
\frac{1}{64\pi^2 \hat{s}} \sum_{\lambda_3,\lambda_4}
\!\!|M_{\lambda_1\lambda_2\lambda_3\lambda_4}|^2 ,
\end{equation}
where $\lambda_1, \lambda_2$ are helicity of the incoming bosons,
and $\lambda_3, \lambda_4$ are helicities of the outgoing photons.
 In
its turn, each of the helicity amplitudes
$M_{\lambda_1\lambda_2\lambda_3\lambda_4}$ is a sum of an anomaly
and SM terms,
\begin{equation}\label{anom+SM}
M_{\lambda_1\lambda_2\lambda_3\lambda_4} =
M_{\lambda_1\lambda_2\lambda_3\lambda_4}^{\mathrm{anom}} +
M_{\lambda_1\lambda_2\lambda_3\lambda_4}^{\mathrm{SM}} \;.
\end{equation}

To obtain constraints directly on the couplings $\zeta_1$ and
$\zeta_2$ in the effective Lagrangian \eqref{Lagrangian}, we assume
that there is no new quartic interactions of neutral gauge bosons
\emph{except for the $Z\gamma\gamma\gamma$ vertex}. For practical
use it means that the anomalous contribution to $d\hat{\sigma}
(V_1V_2\rightarrow \gamma\gamma)$ in \eqref{cs} comes from the VBF
process $Z\gamma\rightarrow\gamma\gamma$ only ($V_1= \gamma, V_2 =
Z$ or $V_1 = Z, V_2 = \gamma$ in Fig.~\ref{fig:mu-VBF-mu_m}). The
explicit expressions for
$M_{\lambda_1\lambda_2\lambda_3\lambda_4}^{\mathrm{anom}}$ are given
in Appendix~A.

As for a SM contribution to the cross section, all VBF processes,
like $\gamma\gamma\rightarrow\gamma\gamma$, $\gamma
Z\rightarrow\gamma\gamma$ ($Z\gamma\rightarrow\gamma\gamma$), and
$ZZ\rightarrow\gamma\gamma$, should be taken into account (see
Fig.~\ref{fig:mu-VBF-mu_m}). Each of the SM amplitudes in
\eqref{anom+SM} is a sum of the fermion and $W$ boson one-loop
amplitudes,
\begin{equation}\label{f+W_ew}
M^{\mathrm{SM}} = M^{\mathrm{SM}}_f + M^{\mathrm{SM}}_W  \;.
\end{equation}
The analytical expressions for the SM helicity amplitudes
$M^{\mathrm{SM}}_f$ and $M^{\mathrm{SM}}_W$ are well-known, see
\cite{Jikia:1994}-\cite{Gounaris:2000}. Note that
$M^{\mathrm{SM}}_W$ dominates over $M^{\mathrm{SM}}_f$ at $\sqrt{s}
> 200$ GeV.

%%%%%%%%%%%%%%%%%%%%%%%%%%%%%%
\section{Numerical analysis} %
%%%%%%%%%%%%%%%%%%%%%%%%%%%%%%

\begin{figure}[htb]
\begin{center}
\includegraphics[scale=0.52]{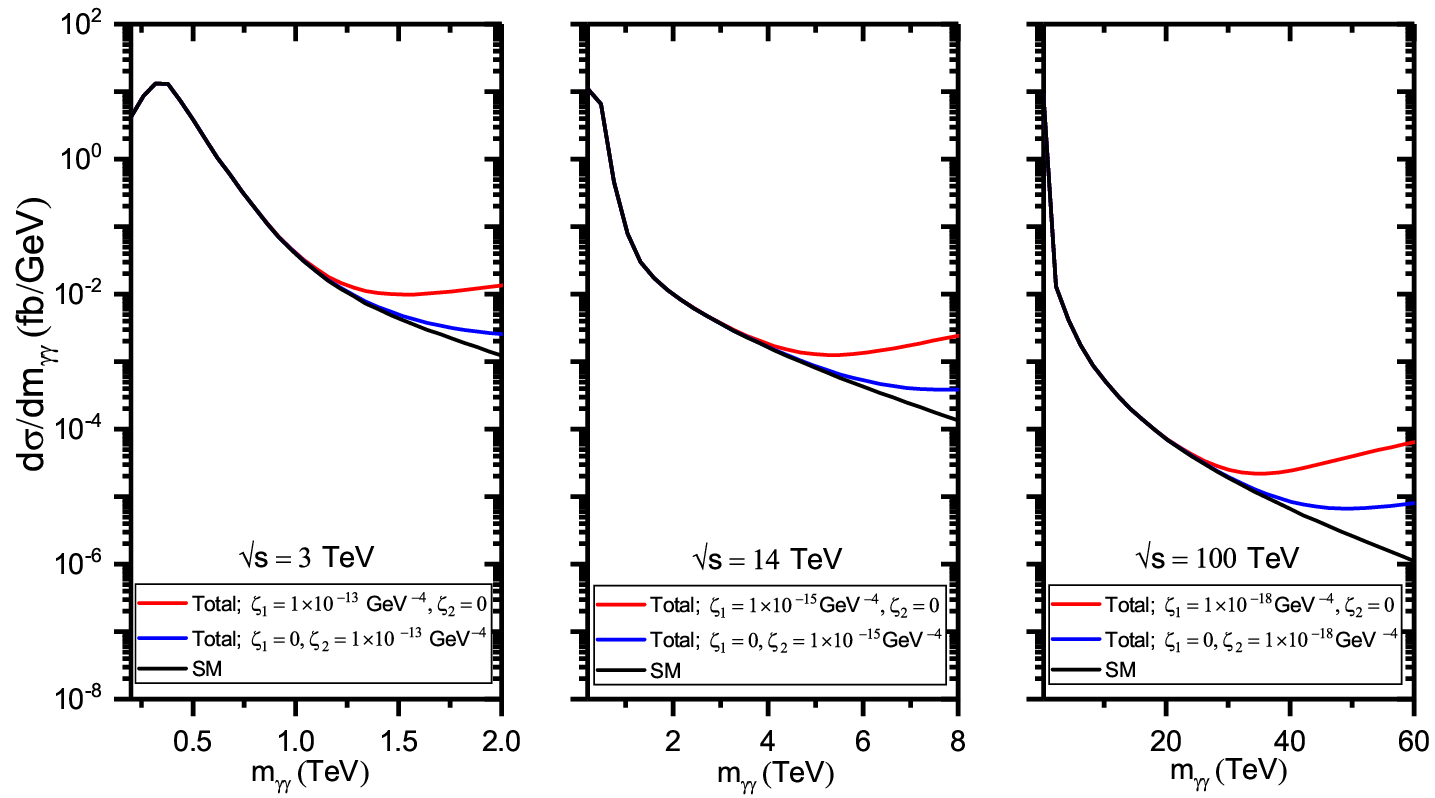}
\caption{The differential cross sections for the  $\mu^+\mu^-
\rightarrow \mu^+ \gamma\gamma \mu^-$ scattering at the future muon
collider versus diphoton invariant mass $m_{\gamma\gamma}$. The
center-of-mass energy is equal to 3 TeV (left panel), 14 TeV (middle
panel), and 100 TeV (right panel).}
\label{fig:WDCS}
\end{center}
\end{figure}

The differential cross sections for the collision energy of
$\sqrt{s} = 3$ TeV, 14 TeV, and 100 TeV are presented in
Fig.~\ref{fig:WDCS} versus diphoton invariant mass
$m_{\gamma\gamma}$. The cuts on the rapidity of the outgoing
photons, $|\eta| < 2.5$, and on the photon transverse momenta,
$p_\bot > 80$ GeV, were applied. For each value of $s$ we present
curves for two sets of the anomalous couplings, ($\zeta_1 \neq 0,
\zeta_2 = 0$) and ($\zeta_1 = 0, \zeta_2 \neq 0$). The SM
contributions to the cross sections are also shown. Because of the
cut \eqref{angle_cut}, the dominant background is the SM process
$\mu^+\mu^- \rightarrow \mu^+ \gamma\gamma \mu^{+}$~\cite{Han:2023}.

\begin{figure}[htb]
\begin{center}
\includegraphics[scale=0.52]{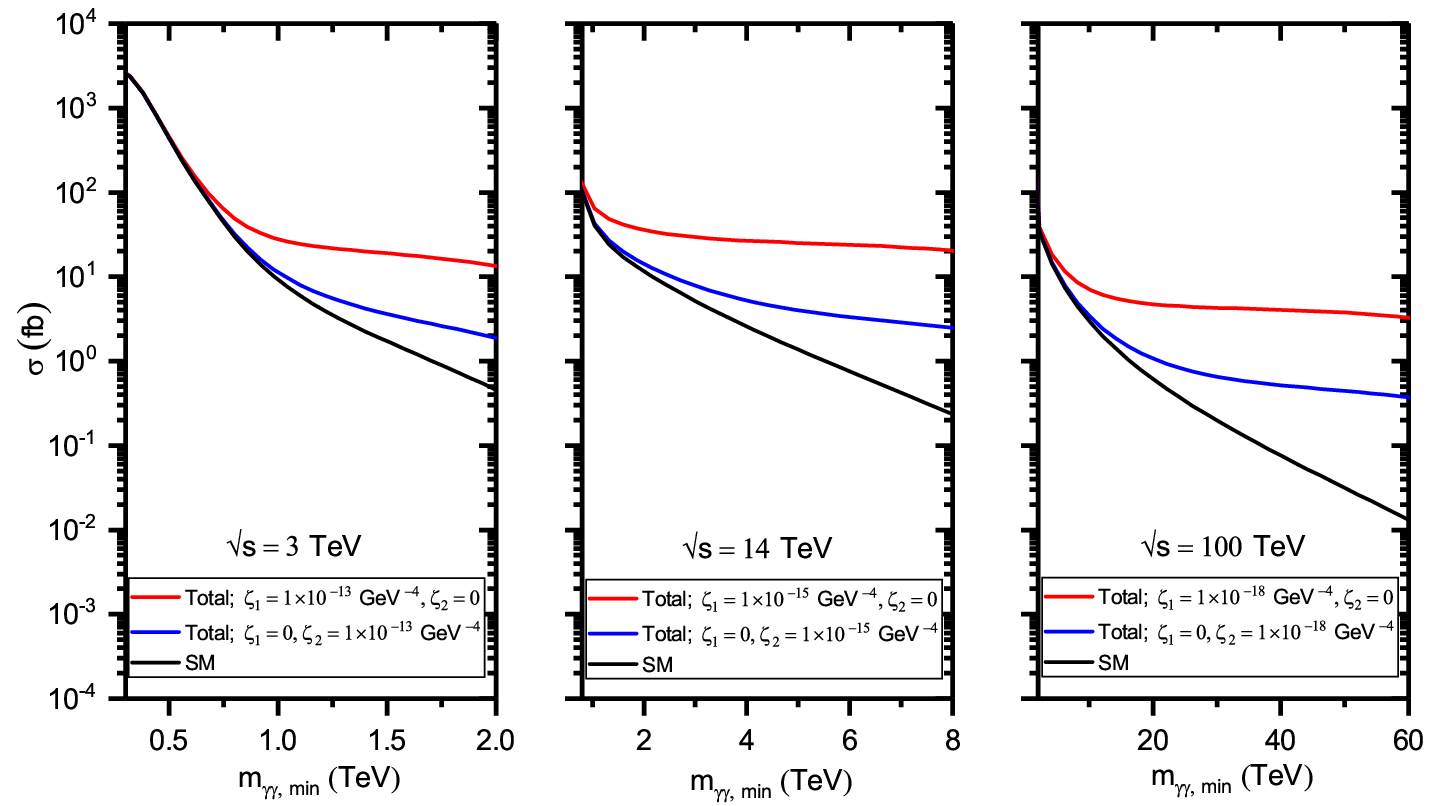}
\caption{The total cross sections $\sigma(m_{\gamma\gamma}
> m_{\gamma\gamma,\min})$ for the $\mu^+\mu^- \rightarrow \mu^+ \gamma\gamma \mu^-$ scattering at
the future muon collider versus minimal value of the invariant mass
of the photon pair $m_{\gamma\gamma}$.}
\label{fig:WCUTCS}
\end{center}
\end{figure}

The total cross sections as functions of the minimal value of the
diphoton invariant mass $m_{\gamma\gamma}$ are given in
Fig.~\ref{fig:WCUTCS}. We see that for $\sqrt{s} = 3$ TeV the total
cross section starts to dominate the SM one at approximately
$m_{\gamma\gamma} \gtrsim \sqrt{s}/2$. For higher collision energies
of 14 TeV and 100 TeV it takes place already if $m_{\gamma\gamma}
\gtrsim \sqrt{s}/7$.
%
%%%%%%%%%%%%%%%%%%%%%%%%%%%%%%%%%%%%%%%%%
% Figure 2. Differential cross sections %
%%%%%%%%%%%%%%%%%%%%%%%%%%%%%%%%%%%%%%%%%
%
%%%%%%%%%%%%%%%%%%%%%%%%%%%%%%%%%%
% Figure 3. Total cross sections %
%%%%%%%%%%%%%%%%%%%%%%%%%%%%%%%%%%

Let $s(b)$ be the total number of signal (background) events, and
$\delta$ the percentage systematic error. Then the exclusion
significance is given by \cite{SS}
\begin{equation}\label{S_excl}
S_{\mathrm{excl}} = \sqrt{ 2\left[ s - b \ln \left( \frac{b + s +
x}{2b} \right) - \frac{1}{\delta^2}\ln \left( \frac{b - s + x}{2b}
\right) - (b + s -x) \left( 1 + \frac{1}{\delta^2 b} \right) \right]
} ,
\end{equation}
with
\begin{equation}\label{x}
x = \sqrt{ (s+b)^2 - \frac{4s\delta^2 b^2}{(1 + \delta^2 b)} } \;.
\end{equation}
We define the regions $S_{\mathrm{excl}} \leqslant 1.645$ as the
regions that can be excluded at the 95\% C.L. To reduce the SM
background, we additionally impose the lower cut on the invariant
mass of the outgoing photons. Namely, we assume that the minimal
value of $m_{\gamma\gamma}$ is equal to 2 TeV, 8 TeV, and 50 TeV for
$\sqrt{s} = 3$ TeV, 14 TeV, and 100 TeV, respectively. Following
\cite{muon_eff_1,muon_eff_2}, we assumed that for $p_\bot > 80$ GeV
(our cut) the muon reconstruction efficiency is larger than $85\%$ .

%%%%%%%%%%%%%%%%%%%%%%%%%%%%%%%%%%%
% Figure 4. Exclusion bound 3 TeV %
%%%%%%%%%%%%%%%%%%%%%%%%%%%%%%%%%%%
\begin{figure}[htb]
\begin{center}
\includegraphics[scale=0.6]{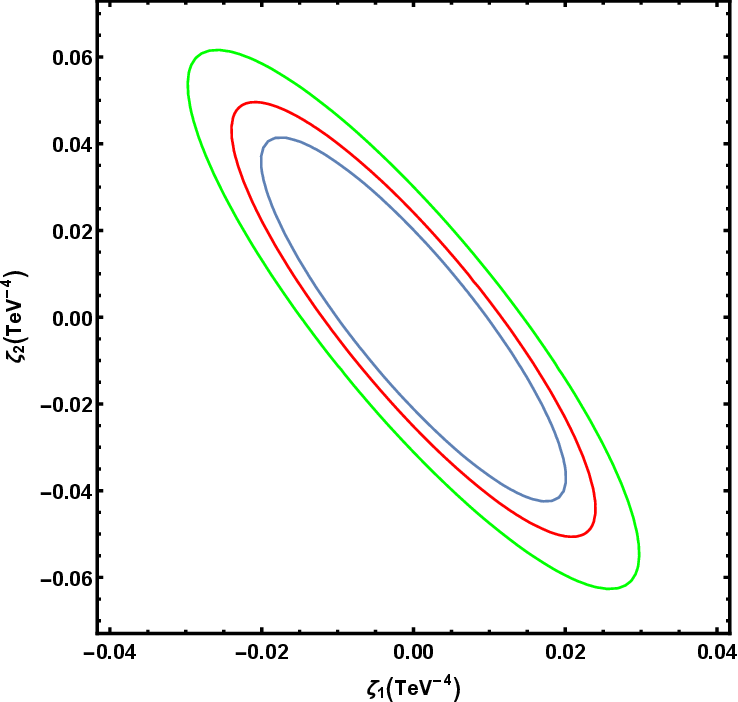}
\caption{The 95\% C.L. exclusion regions for the couplings $\zeta_1,
\zeta_2$ in the unpolarized $\mu^+\mu^- \rightarrow
\mu^+\gamma\gamma \mu^-$ scattering at the future muon collider with
the systematic errors $\delta = 0\%$ (blue ellipse), $\delta = 5\%$
(red ellipse), and $\delta = 10\%$ (green ellipse). The inner
regions of the ellipses are inaccessible. The collision energy is
$\sqrt{s} = 3$ TeV, the integrated luminosity is $L = 1$ ab$^{-1}$.
The cut on the diphoton invariant mass $m_{\gamma\gamma} > 2$ TeV is
imposed.}
\label{fig:excl_1500}
\end{center}
\end{figure}
%
%%%%%%%%%%%%%%%%%%%%%%%%%%%%%%%%%%%%
% Figure 5. Exclusion bound 14 TeV %
%%%%%%%%%%%%%%%%%%%%%%%%%%%%%%%%%%%%
\begin{figure}[htb]
\begin{center}
\includegraphics[scale=0.6]{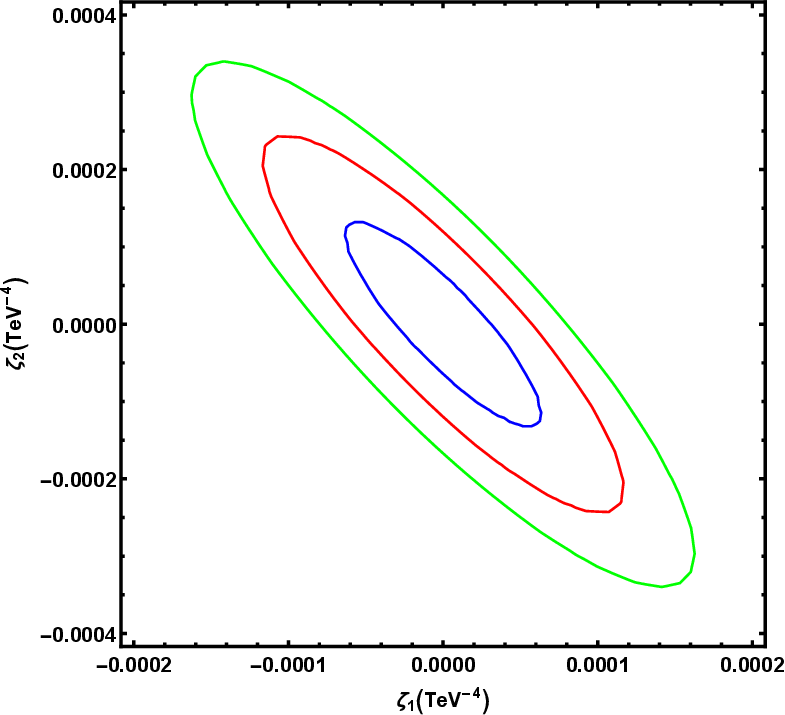}
\caption{The same as in Fig.~\ref{fig:excl_1500}, but for $\sqrt{s}
= 14$ TeV and $L = 20$ ab$^{-1}$. The cut $m_{\gamma\gamma} > 8$ TeV
is used.}
\label{fig:excl_7000}
\end{center}
\end{figure}
%
%%%%%%%%%%%%%%%%%%%%%%%%%%%%%%%%%%%%%
% Figure 6. Exclusion bound 100 TeV %
%%%%%%%%%%%%%%%%%%%%%%%%%%%%%%%%%%%%%
\begin{figure}[htb]
\begin{center}
\includegraphics[scale=0.6]{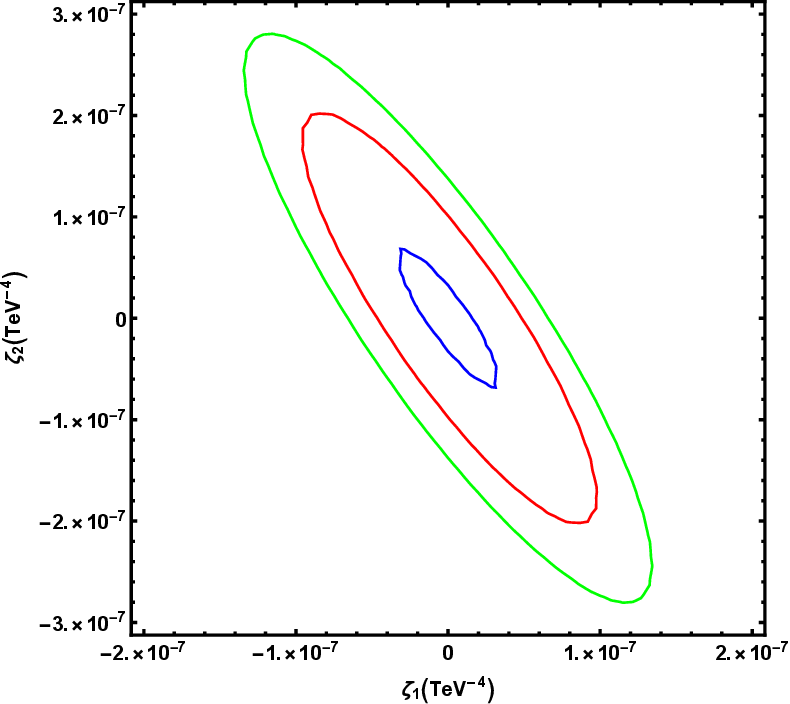}
\caption{The same as in Fig.~\ref{fig:excl_1500}, but for $\sqrt{s}
= 100$ TeV and $L = 1000$ ab$^{-1}$. The cut $m_{\gamma\gamma} > 50$
TeV is imposed.}
\label{fig:excl_50000}
\end{center}
\end{figure}

Our 95\% C.L. exclusion regions for AQGCs for the unpolarized
process $\mu^+\mu^- \rightarrow \mu^+ \gamma\gamma \mu^-$ at the
muon collider are shown in
Figs.~\ref{fig:excl_1500}-\ref{fig:excl_50000}. Note that in the
energy region $s \gg m_Z^2$ the squared amplitude is proportional to
the coupling combination $48\zeta_1^2 + 11\zeta_1\zeta_2 +
40\zeta_2^2$. In such a case, the exclusion regions are ellipses in
the plane $\zeta_1 - \zeta_2$, as we obtained. The best bounds on
AQGCs when one of the couplings $\zeta_1, \zeta_2$ is taken to be
zero are given in Tab.~\ref{tab:excl_best_values}.

%%%%%%%%%%%
% Table 1 %
%%%%%%%%%%%
\begin{table}[h]
    \centering \caption{The 95\% C.L. exclusion bounds on the anomalous
        quartic couplings $\zeta_1$ and $\zeta_2$, with the integrated luminosity of 1 ab$^{-1}$,
        20 ab$^{-1}$, and 1000 ab$^{-1}$ for the 3 TeV, 14 TeV, and 100 TeV muon collider.
        \bigskip} \label{tab:excl_best_values}
    \begin{tabular}{||c||c|c|c|c||}
        \hline
        & & 3 TeV & 14 TeV & 100 TeV
        \\
        \hline
        \makecell{$|\zeta_1|$, TeV$^{-4}$ \\ $(\zeta_2=0)$} & \makecell{$\delta=0\%$ \\  $\delta=5\%$ \\ $\ \, \delta=10\%$ } &
        $\makecell{ 9.90\times10^{-3} \\ 1.18\times10^{-2} \\ 1.46\times10^{-2}}$ &
        $\makecell{ 3.13\times10^{-5} \\ 5.75\times10^{-5} \\ 8.01\times10^{-5}}$ &
        $\makecell{ 1.64\times10^{-8} \\ 4.79\times10^{-8} \\ 6.61\times10^{-8}}$ \\
        \hline
        \makecell{$|\zeta_2|$, TeV$^{-4}$ \\  $(\zeta_1=0)$} & \makecell{$\delta=0\%$ \\ $\delta=5\%$ \\ $\ \, \delta=10\%$ } &
        $\makecell{ 2.07\times10^{-2} \\ 2.46\times10^{-2} \\ 3.06\times10^{-2}}$ &
        $\makecell{ 6.54\times10^{-5} \\ 1.53\times10^{-4} \\ 1.67\times10^{-4}}$ &
        $\makecell{ 3.43\times10^{-8} \\ 1.02\times10^{-7} \\ 1.38\times10^{-7}}$ \\
        \hline
    \end{tabular}
\end{table}

%%%%%%%%%%%%%%%%%%%%%%%%%%%%%%%%%%%%%%%%%%%%%%%%%%%
\section{Unitarity bounds on anomalous couplings} %
%%%%%%%%%%%%%%%%%%%%%%%%%%%%%%%%%%%%%%%%%%%%%%%%%%%

Now let us derive constraints imposed by partial-wave unitarity. To
do it, we can use the partial-wave expansion of the helicity
amplitude in the center-of-mass system \cite{Jacob:2000}
\begin{align}\label{helicity_ampl_expansion}
M_{\lambda_1\lambda_2\lambda_3\lambda_4}(s, \theta, \varphi) &=
16\pi \sum_J (2J + 1) \sqrt{(1 + \delta_{\lambda_1\lambda_2})(1 +
\delta_{\lambda_3\lambda_4})}
\nonumber \\
&\times \,e^{i(\lambda - \mu)\phi} \,d^J_{\lambda\mu}(\theta)
\,T^J_{\lambda_1\lambda_2\lambda_3\lambda_4}(s) \;,
\end{align}
where $\lambda = \lambda_1 - \lambda_2$, $\mu = \lambda_3 -
\lambda_4$, $\theta(\phi)$ is the polar (azimuth) scattering angle.
If we choose the plane $(x - z)$ as a scattering plane, then $\phi =
0$ in \eqref{helicity_ampl_expansion}. Here
$d^J_{\lambda\mu}(\theta)$ is the Wigner (small) $d$-function
\cite{Wigner}. In particular, we have \cite{I_K:2021_2}
\begin{equation}\label{d-function_00}
d^J_{00}(z) = P_J(z) \;,
\end{equation}
and
\begin{equation}\label{d-function_22}
d^J_{2 \pm2}(z) = (\pm 1)^J \!\left( \frac{1 \pm z}{2}\right)^{\!2}
\!{}_2F_1 \!\left( 2-J, J+3; 1; \frac{1 \mp z}{2} \right) ,
\end{equation}
where $z = \cos\theta$, $P_J(z)$ is the Legendre polynomial, and
$_2F_1(a,b;c;x)$ is the hypergeometric function. It is enough to
calculate unitarity bounds on our anomalous couplings.

Partial-wave unitarity requires that
\begin{equation}\label{parity_wave_unitarity}
\left| T^J_{\lambda_1\lambda_2\lambda_3\lambda_4}(s) \right| \leq 1
\;.
\end{equation}
Note that $T^J_{\lambda_1\lambda_2\lambda_3\lambda_4}(s)$ is related
to the partial-wave amplitude with the same helicities by formula
\begin{align}\label{parity wave_func}
T^J_{\lambda_1\lambda_2\lambda_3\lambda_4}(s) &= \frac{1}{32\pi}
\frac{1}{\sqrt{(1 + \delta_{\lambda_1\lambda_2})(1 +
\delta_{\lambda_3\lambda_4})}} \int\limits_{-1}^1 \!\!
M_{\lambda_1\lambda_2\lambda_3\lambda_4}(s, z)
\,d^{J}_{\lambda\mu}(z) \,dz \;.
\end{align}
The anomalous helicity amplitudes
$M_{\lambda_1\lambda_2\lambda_3\lambda_4}$ for process
\eqref{process} are calculated in Appendix~A.

In the limit $s \gg m_Z^2$ the most stringent bounds come from the
helicity amplitudes $M_{++++}$, $M_{++--}$, $M_{+-+-}$, and
$M_{+--+}$ (as well as from $M_{-+++}$, $M_{-+--}$, $M_{--+-}$, and
$M_{---+}$). On the other hand, these amplitudes coincide with
$\gamma\gamma\gamma\gamma$ anomalous amplitudes up to small
corrections $\mathrm{O(m_Z^2/s)}$, see, for instance, Appendix in
\cite{I_K:2023}. It means that unitarity bounds derived in
\cite{I_K:2023} are also valid for our anomalous couplings. They
look like
\begin{equation}\label{unitariry_bounds}
|\zeta_1| < 2\pi s^{-2} \;, \quad |\zeta_2| \leq (16\pi/3) s^{-2}
\;.
\end{equation}
As we see, the unitarity demands that the coupling $|\zeta_1|$ must
be less than $7.8 \times 10^{-2}$ TeV$^{-4}$, $1.64 \times 10^{-4}$
TeV$^{-4}$, and $6.28 \times 10^{-8}$ TeV$^{-4}$ for $\sqrt{s} = 3$
TeV, 14 TeV, and 100 TeV, respectively. Correspondingly, the
unitarity bounds for the coupling $|\zeta_2|$ are $0.21$ TeV$^{-4}$,
$4.36 \times 10^{-4}$ TeV$^{-4}$, and $1.68 \times 10^{-7}$
TeV$^{-4}$. These constraints should be compared with the bounds on
AQGCs presented in Tab.~\ref{tab:excl_best_values}. Let us note that
in eq.~\eqref{unitariry_bounds} one actually has to use the
invariant energy square $\hat{s}$ of the subprocess $V_1 V_2
\rightarrow \gamma\gamma$ instead of the collision energy $s$. To
conclude, the unitarity is not violated in the region of AQGCs
studied in the present paper.

%%%%%%%%%%%%%%%%%%%%%%%
\section{Conclusions} %
%%%%%%%%%%%%%%%%%%%%%%%

In the present paper we have probed the quartic gauge couplings
(QGCs) of the anomalous $\gamma\gamma\gamma Z$ interaction in the
$\mu^+\mu^- \rightarrow \mu^+ \gamma\gamma \mu^-$ scattering at the
high energy muon collider with unpolarized beams. The sum over the
photon polarization states are also assumed. Our process is the
exclusive one by requiring the outgoing muons to be observable in
the detector range $10^\circ < \theta < 170^\circ$. The collision
energy of $\sqrt{s} = 3$ TeV, 14 TeV, and 100 TeV with the
integrated luminosity of $L = 1$ ab$^{-1}$, 20 ab$^{-1}$, and 1000
ab$^{-1}$, respectively, are addressed. Both differential and total
cross sections are calculated depending on the invariant mass of the
diphoton system $m_{\gamma\gamma}$. We have used the cut on the
photon rapidity, $|\eta| < 2.5$. The cut on the photon transverse
momenta, $p_\bot > 80$ GeV, has been also imposed. The 95\% C.L.
exclusion regions for the AQGCs $\zeta_1$ and $\zeta_2$ are
calculated. They are ellipses in the $\zeta_1 - \zeta_2$ plane, see
Figs.~\ref{fig:excl_1500}-\ref{fig:excl_50000}. The limits on
$\zeta_1$ (for $\zeta_2 = 0$) and $\zeta_2$ (for $\zeta_2 = 0$) are
collected in Tab.~1 for three values of the systematic error (0\%,
5\%, and 10\%). We have investigated the constraints imposed by the
partial-wave unitarity. It has been shown that it is not violated
for the AQGCs values examine in our paper.

As was already mentioned in Introduction, bounds on the quartic
anomalous couplings were obtained for future hadron colliders
\cite{Senol:2022_1}-\cite{Senol:2022}. In particular, the 95\% C.L.
limits on AQGCs for the LHC an FCC-hh colliders are collected in
Tab.~IV in \cite{Senol:2022}. One can see that for the 14 TeV HL-LHC
with the integrated luminosity of $L = 3$ ab$^{-1}$ the best 95\%
C.L. limit on AQGCs is equal to $2.1 \times 10^{-1}$ TeV$^{-4}$. The
most stringent bound for the HE-LHC with the center-of-mass energy
$\sqrt{s} = 27$ TeV and $L = 15$ ab$^{-1}$ is obtained to be $2.8
\times 10^{-2}$ TeV$^{-4}$. Finally, limits on AQGCs for the future
FCC-hh collider are derived in \cite{Senol:2022}, the best one is
equal to $5.4 \times 10^{-4}$ TeV$^{-4}$. To compare, our best bound
on the coupling $\zeta_1$ ($\zeta_2$) for the 14 TeV muon collider
with $L = 20$ ab$^{-1}$ is equal to 3.13 (6.54) $\times 10^{-5}$
TeV$^{-4}$. All this demonstrates a great potential of the future
muon collider in probing anomalous interactions of the neutral gauge
bosons.

%%%%%%%%%%%%%%%%%%%%%%%%%%%%%%%%%%%%%%%%%%%%%%%%%%%%%%%%%%%%%%%%%%%%%

%%%%%%%%%%%%%%
% Appendix A %
%%%%%%%%%%%%%%

\setcounter{equation}{0}
\renewcommand{\theequation}{A.\arabic{equation}}

\section*{Appendix A. Anomalous helicity amplitudes}

First consider helicity amplitudes
$\bar{M}_{\lambda_1\lambda_2\lambda_3\lambda_4}$ of the process with
an \emph{outgoing} $Z$ boson,
\begin{equation}\label{process_Z_incoming}
\gamma(p_1,\lambda_1) \gamma(p_2,\lambda_2) \rightarrow
\gamma(p_3,\lambda_3) Z(p_4,\lambda_4) \;.
\end{equation}
They can be obtained from explicit expressions given in our paper
\cite{I_K:2021_2}. There the helicity amplitudes
$\tilde{M}_{\lambda_1\lambda_2\lambda_3\lambda_4}$ were derived from
the Lagrangian
\begin{equation}\label{Lagrangian_IK}
\mathcal{L}_{\gamma\gamma\gamma Z} = g_1 F^{\rho\mu} F^{\alpha\nu}
\partial_\rho F_{\mu\nu} Z_\alpha + g_2 F^{\rho\mu} F^\nu_\mu \partial_\rho F_{\alpha\nu}
Z^\alpha  \;.
\end{equation}
They look like
\begin{equation}\label{helicity_ampl_IK}
\tilde{M}_{\lambda_1\lambda_2\lambda_3\lambda_4} = g_1
\tilde{M}_{\lambda_1\lambda_2\lambda_3\lambda_4}^{(1)} + g_2
\tilde{M}_{\lambda_1\lambda_2\lambda_3\lambda_4}^{(2)} \;,
\end{equation}
with the explisit expressions for
$\tilde{M}_{\lambda_1\lambda_2\lambda_3\lambda_4}^{(i)}$ ($i=1,2$)
given in Appendix~B of \cite{I_K:2021_2}. Then, according to
eq.~(2.7) in \cite{I_K:2021_2}, the helicity amplitude
$\bar{M}_{\lambda_1\lambda_2\lambda_3\lambda_4}$ corresponding to
the effective Lagrangian \eqref{Lagrangian} is defined as
\begin{equation}\label{helicity_amlitudes_relation}
\bar{M}_{\lambda_1\lambda_2\lambda_3\lambda_4} = -2(4\zeta_1 +
\zeta_2) \tilde{M}_{\lambda_1\lambda_2\lambda_3\lambda_4}^{(1)} +
2\zeta_2 \tilde{M}_{\lambda_1\lambda_2\lambda_3\lambda_4}^{(2)} \;.
\end{equation}
Bose-Einstein statistics and parity invariance demand that there
exist six independent helicity amplitudes
$\bar{M}_{\lambda_1\lambda_2\lambda_3\lambda_4}$ with $\lambda_1 =
+1$ and $\lambda_4 = \pm 1$
\begin{align}\label{independent_helicity_ampl}
\bar{M}_{++++}(s,t,u) &= -\frac{(4\zeta_1 + 3\zeta_2)}{2} \,s(u+t)
\;,
\nonumber \\
\bar{M}_{+++-}(s,t,u) &= 0 \;,
\nonumber \\
\bar{M}_{++-+ }(s,t,u) &= -(4\zeta_1 + \zeta_2) \frac{tu}{t+u} m_Z^2
\;,
\nonumber \\
\bar{M}_{+-++}(s,t,u) &= -\frac{(4\zeta_1 + 3\zeta_2)}{2}
\frac{tu}{t+u} m_Z^2 \;,
\nonumber \\
\bar{M}_{++--}(s,t,u) &= -(4\zeta_1 + \zeta_2) \frac{s(t^2 + tu +
u^2)}{t+u} \;,
\nonumber \\
\bar{M}_{+-+-}(s,t,u) &= -\frac{(4\zeta_1 + 3\zeta_2)}{2}
\,\frac{su^2}{t+u} \;.
\end{align}
where $s$, $t$, $u$ are Mandelstam variables, and $s+t+u = m_Z^2$.
We also have the crossing relations
\begin{align}\label{dependent_helicity_ampl}
\bar{M}_{+--+}(s,t,u) &= \bar{M}_{+-+-}(s,u,t) = -\frac{(4\zeta_1 +
3\zeta_2)}{2} \,\frac{st^2}{t+u} \;,
\nonumber \\
\bar{M}_{+---}(s,t,u) &= \bar{M}_{+-++}(s,u,t) = -\frac{(4\zeta_1 +
3\zeta_2)}{2} \frac{tu}{t+u} m_Z^2 \;.
\end{align}
Equations \eqref{independent_helicity_ampl},
\eqref{dependent_helicity_ampl} represent eight helicity amplitudes
with $\lambda_1 = +1$ and $\lambda_4 = \pm 1$. For $\lambda_1 = +1$,
$\lambda_4 = 0$ we get
\begin{align}\label{ingoing_helicity_ampl_0}
\bar{M}_{+++0}(s,t,u) &= 0 \;,
\nonumber \\
\bar{M}_{++-0}(s,t,u) &= - \frac{i(4\zeta_1 + \zeta_2)}{\sqrt{2}}
\sqrt{stu} \,\frac{t-u}{t+u} m_Z \;,
\nonumber \\
\bar{M}_{+-+0}(s,t,u) &= \frac{i(4\zeta_1 + 3\zeta_2)}{\sqrt{2}}
\frac{u\sqrt{stu}}{t+u} m_Z \;,
\nonumber \\
\bar{M}_{+--0}(s,t,u) &= \bar{M}_{+-+0}(s,u,t) = \frac{i(4\zeta_1 +
3\zeta_2)}{\sqrt{2}} \frac{t\sqrt{stu}}{t+u} m_Z \;.
\end{align}
The helicity amplitudes with $\lambda_1 = -1$ can be obtained from
them by use of parity relation,
\begin{equation}\label{parity_relations_1}
\bar{M}_{-\lambda_1\lambda_2\lambda_3\lambda_4}(s,t,u) =
(-1)^{1-\lambda_4}
\bar{M}_{\lambda_1-\lambda_2-\lambda_3-\lambda_4}(s,t,u) \;.
\end{equation}

Now consider the process with an \emph{incoming} $Z$ boson,
\begin{equation}\label{process_Z_outgoing}
\gamma(p_1,\lambda_1) Z(p_2,\lambda_2) \rightarrow
\gamma(p_3,\lambda_3) \gamma(p_4,\lambda_4) \;.
\end{equation}
If time-inversion invariance holds, we have the relation
\begin{equation}\label{time_inversion_relations}
M_{\lambda_1\lambda_2\lambda_3\lambda_4}(s,t,u) =
\bar{M}_{\lambda_3\lambda_4\lambda_1\lambda_2}(s,t,u) \;.
\end{equation}
Then using eqs.~\eqref{independent_helicity_ampl} and
\eqref{ingoing_helicity_ampl_0} we find that the helicity amplitudes
of the process \eqref{process_Z_outgoing} are given by
\begin{align}\label{helicity_ampl_Z_outgoing}
M_{++++}(s,t,u) &= \bar{M}_{++++}(s,t,u) = -\frac{(4\zeta_1 +
3\zeta_2)}{2} \,s(t+u) \;,
\nonumber \\
M_{+++-}(s,t,u) &= \bar{M}_{+-++}(s,u,t) = -\frac{(4\zeta_1 +
3\zeta_2)}{2} \frac{tu}{t+u} m_Z^2 \;,
\nonumber \\
M_{++-+ }(s,t,u) &= \bar{M}_{+---}(s,t,u) = -\frac{(4\zeta_1 +
3\zeta_2)}{2} \frac{tu}{t+u} m_Z^2 \;,
\nonumber \\
M_{+-++}(s,t,u) &= \bar{M}_{+++-}(s,t,u) = 0 \;,
\nonumber \\
M_{++--}(s,t,u) &= \bar{M}_{++--}(s,t,u) = -(4\zeta_1 + \zeta_2)
\,\frac{s(t^2 + tu + u^2)}{t+u} \;,
\nonumber \\
M_{+-+-}(s,t,u) &=  \bar{M}_{+-+-}(s,t,u) = -\frac{(4\zeta_1 +
3\zeta_2)}{2} \,\frac{su^2}{t+u} \;,
\nonumber \\
M_{+--+}(s,t,u) &= \bar{M}_{+--+}(s,t,u) = -\frac{(4\zeta_1 +
3\zeta_2)}{2} \,\frac{st^2}{t+u} \;,
\nonumber \\
M_{+---}(s,t,u) &= \bar{M}_{++-+ }(s,t,u) = -(4\zeta_1 + \zeta_2)
\frac{tu}{t+u} m_Z^2 \;,
\end{align}
and
\begin{align}\label{helicity_ampl_Z_outgoing_0}
M_{+0++}(s,t,u) &= \bar{M}_{+++0}(s,t,u) = 0 \;,
\nonumber \\
M_{+0+-}(s,t,u) &= \bar{M}_{+-+0}(s,t,u) = \frac{i(4\zeta_1 +
3\zeta_2)}{\sqrt{2}} \frac{u\sqrt{stu}}{t+u} m_Z \;,
\nonumber \\
M_{+0-+}(s,t,u) &= -\bar{M}_{+--0}(s,t,u) = - \frac{i(4\zeta_1 +
3\zeta_2)}{\sqrt{2}} \frac{t\sqrt{stu}}{t+u} m_Z \;,
\nonumber \\
M_{+0--}(s,t,u) &= -\bar{M}_{++-0}(s,t,u)  = \frac{i(4\zeta_1 +
\zeta_2)}{\sqrt{2}} \sqrt{stu} \,\frac{t-u}{t+u} m_Z \;.
\end{align}
The helicity amplitudes $M_{\lambda_1\lambda_2\lambda_3\lambda_4}$
with $\lambda_1 = -1$ can be obtained from
\eqref{helicity_ampl_Z_outgoing}, \eqref{helicity_ampl_Z_outgoing_0}
by use of parity relation
\begin{equation}\label{parity_relations}
M_{-\lambda_1\lambda_2\lambda_3\lambda_4}(s,t,u) =
(-1)^{1-\lambda_2}
M_{\lambda_1-\lambda_2-\lambda_3-\lambda_4}(s,t,u) \;.
\end{equation}
The amplitudes
\eqref{helicity_ampl_Z_outgoing}-\eqref{parity_relations} will be
used for our numerical calculations.

%%%%%%%%%%%%%%%%%%%%%%%%%%%%%%%%%%%%%%%%%%%%%%%%%%%%%%%%%%%%%%%%%%%%%

%%%%%%%%%%%%%%
% References %
%%%%%%%%%%%%%%

%%%%%%%%%%%%%%%%%%%%%%%%%%%%%%%%%%%%%%%%%%%%%%%%%%%%%%%%%%%%%%%%%%%%

%%%%%%%%%%%%%%%
\end{document}